# Photonic Time Stretch Enhanced Recording Scope

Shalabh Gupta and Bahram Jalali

UCLA, LOS ANGELES, CA 90095-1594

*Abstract*—We demonstrate a new mode of operation for the Time Stretched A/D converter. The technique is superior to the sampling scope in that it permits the capture of non periodic events such as clustered noise.

## I. INTRODUCTION

With speed of electronics and communication data rates moving up, the demand for higher performance digitizers has been ever increasing. Very high bandwidth digitizers are required in radar, communication and scientific instruments. In an oscilloscope, high speed digitizers are also used for measuring eye diagrams for characterization of high speed serial links and in fault diagnostic methods such as time-domain reflectometry (TDR).

Real-time digitizers can continuously sample and digitize signals for long durations but have limited bandwidths. On the other hand, equivalent-time digitizers (known as sampling scopes) rely on repetitive or clock synchronous nature of the signals to reconstruct them in time. In sampling oscilloscopes, the signal is sampled at slow clock edges (at MHz frequencies) and reconstructed digitally, requiring a long time period to obtain the original signal with high fidelity. Equivalent time sampling is similar to the strobe light technique used for measuring cyclical events which are much faster than the speed of the detector. For instance, when strobe light is flashed on a vibrating tuning fork at a desirable frequency, it can look stationary or very slowly vibrating. In this paper, we demonstrate the use of photonic Time-Stretch ADC [1] in a similar way to obtain the TiSER (Time-Stretch Enhanced Recording) scope. The differences with the traditional sampling scope are two fold: (1) instead of a single sample, a time segment is captured at every clock cycle, and (2) the segment is slowed down in analog domain reducing the bandwidth requirement of the quantizer. The difference with previous implementation of the time stretch ADC is that the new mode of operation avoids the need for multiple parallel channels [2] when capturing a continuous-time signal.

## II. PHOTONIC TIME STERTCH ADC

In a Time-Stretch ADC (TS-ADC), the effective bandwidth of the electrical signal is reduced prior to digitization by stretching the waveform in time, as shown in Figure 1. As a result, the signal can be recorded by a low cost real-time electronic ADC (analog to digital converter). To accomplish this, the signal is modulated on a pulse of linearly chirped optical carrier obtained from a femto-second mode-locked laser (MLL) or a supercontinuum source. Propagation through dispersive fiber stretches the pulse in time. The photo-detector converts this optical signal back to electrical domain and the resultant RF signal is a time-stretched replica with a much reduced analog bandwidth. Electrical signals up to 90 GHz have been digitized after achieving stretch factor of 250 using the TS-ADC [3]. Fundamentally, bandwidth of the Mach-Zehnder modulator limits the overall system bandwidth.

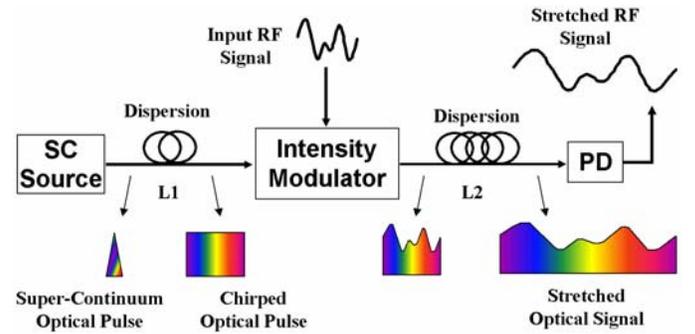

Fig. 1. Schematic diagram of a photonic time stretch preprocessor.

## III. TiSER SCOPE

In the TiSER scope, segments of the signal, each spanning several samples, are captured asynchronously relative to the signal. Each segment is captured by single chirped laser pulse. For the $n^{th}$ laser pulse, the signal recorded at absolute time instant *t* is displayed on time axis of the scope window at time point given by

$$t_{out} = \left( n.T_{laser} + \frac{t - n.T_{laser}}{M} \right) \% \; T_{trigger} \qquad (1)$$

Here, $T_{laser}$ is the pulse repetition period of the laser, $T_{trigger}$ is the trigger period which is synchronous with the signal, M is the stretch factor and % symbol stands for the remainder operation. The trigger and laser periods can easily be obtained in software or using digital phase locking. Since the mode-locked laser (MLL) is asynchronous with the data clock, the optical pulses walk off with respect to data transitions, scanning the data eye at different positions. Though, eye diagrams have previously been captured using TS-ADC [4], this is the first time it is done signals independent of the laser clock with high fidelity.

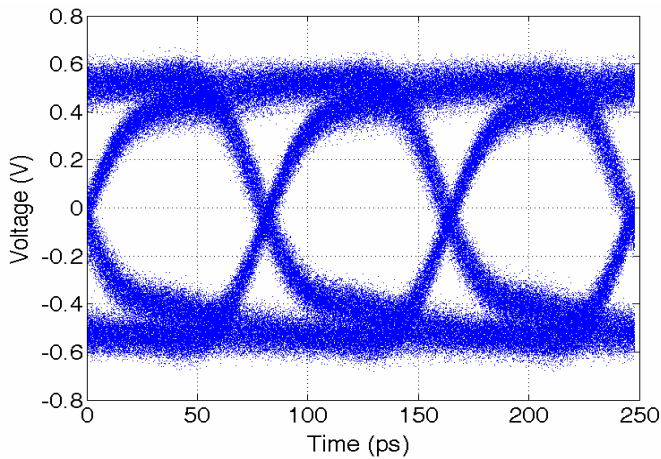

Fig. 2. Eye diagram of 12.1Gbps PRBS data captured using synchronous TS-ADC with stretch factor of 14. The experiment used a 10Gs/s commercial oscilloscope with 4GHz analog bandwidth. However, a 1.5GHz filter was applied digitally to emulate it as a monolithic off-the-shelf ADC.

For the proof of principle, we demonstrate the capture of a 12.1Gbps pseudo-random bit sequence (PRBS) data eye, as shown in Figure 2. The equivalent time oscilloscopes also plot eye diagrams using the same approach. However, this technique can capture signals at much faster rates compared to a sampling oscilloscope and can also show the ultra fast dynamics of rare (non-repetitive) events that happen in the snapshots, as evident from Figure 3.

Use of TiSER for sampling the digital data has several advantages over conventional sampling. Firstly, while a sampling oscilloscope captures only one data point for every sampling strobe, time stretch digitizer can capture multiple independent samples for every laser pulse. In our experiments, 1480 independent samples were captured in every microsecond compared to the highest known sampling frequency 10 MHz in commercially available sampling oscilloscopes, giving 148x higher acquisition speed. Also, as shown in Figure 3, a sampling oscilloscope lacks the capability of capturing any rare event because of very sparse sampling. Even if a rare event is sampled, hardly any information can be gathered from it because connecting samples are missing.

In serial data analysis, jitter is statistically analyzed by obtaining zero crossing points from a sampling scope. However, this process is very slow, as only a very small fraction of samples recorded by the scope lie close to the zero crossings and can be included in the jitter histogram. On the other hand, TiSER samples are connected with each other and all zero crossings occurring in a time stretch pulse can easily be obtained by simple polynomial interpolation. This improves the acquisition times by 3 to 4 orders of magnitudes when compared to today's sampling oscilloscopes, even though samples are captured only at about 100x higher rates. Time stretching holds yet another advantage. While electronic sampling can only give best jitter values of about 200 fs, the jitter added due to sampling clocks in TiSER can be much lower as the electronic jitter scales down by the stretch factor [2]. In TiSER, the dominant source is the laser jitter which can be as low as 18-fs [5], which is an order of magnitude better.

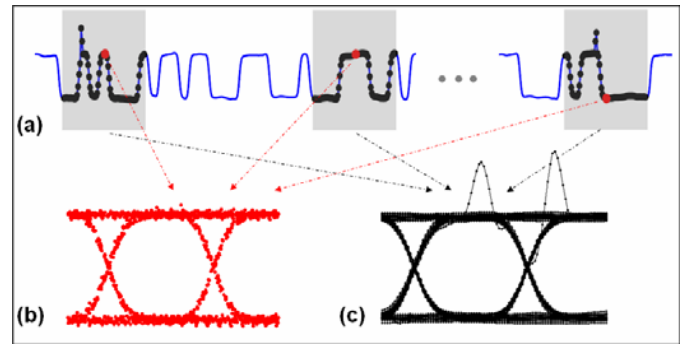

Fig. 3. Comparison between different digitizing mechanisms: (a) Serial digital data waveform (in blue) can be captured by real-time digitizers only for low data rates. (b) Data eye captured by a sampling-oscilloscope (c) Data eye captured by TiSER scope.

## IV. CONCLUSION

We have described a new type of oscilloscope that combines advantages of both real-time digitizers and sampling scopes. As a proof of principle, high fidelity capture of 12.1 Gbps pseudorandom data is demonstrated.

## ACKNOWLEDGEMENT

This work was supported by DARPA-MTO under the PHOBIAC program.